\documentclass[a4paper,11pt]{article}
\usepackage{jheppub}
\usepackage{mathrsfs}
\usepackage{amsfonts}
\usepackage{setspace}
\usepackage{cellspace}
\usepackage{amsmath,amssymb,bm}
\usepackage[colorlinks=true,linkcolor=blue]{hyperref}
\usepackage{xcolor}
\usepackage{epsfig}
\usepackage{slashed}
\usepackage{caption}
\usepackage{hhline,multirow,tabularx}  % for nicer tables
\usepackage{dcolumn}    % align table columns on decimal point
\usepackage{url}        % for URL addresses
\usepackage{braket} 
\definecolor{cyan}{rgb}{0.0, 1.0, 1.0}
\definecolor{applegreen}{rgb}{0.55, 0.71, 0.0}
\definecolor{arylideyellow}{rgb}{0.91, 0.84, 0.42}
\definecolor{bananayellow}{rgb}{1.0, 0.88, 0.21}
\definecolor{burlywood}{rgb}{0.87, 0.72, 0.53}
\definecolor{buff}{rgb}{0.94, 0.86, 0.51}
\definecolor{blond}{rgb}{0.98, 0.94, 0.75}
\definecolor{bisque}{rgb}{1.0, 0.89, 0.77}
\definecolor{bananamania}{rgb}{0.98, 0.91, 0.71}
\definecolor{apricot}{rgb}{0.98, 0.81, 0.69}
\definecolor{almond}{rgb}{0.94, 0.87, 0.8}

\title{Analytic asymptotics of the integrable XXX critical spin chains in the small external-field limit from TBA}
\author[a]{Yizhuang Liu}

\affiliation[a]{Institute of Theoretical Physics,
Jagiellonian University, 30-348 Kraków, Poland}

\emailAdd{yizhuang.liu@uj.edu.pl}

\abstract{In this note, inspired partly by the works on the analytic trans-series of free-energies in 2D integrable QFTs in the UV limit, we study the corresponding asymptotic expansions in the IR limit of the integrable XXX spin chains, at the anti-ferromagnetic critical point. Starting from the linear TBA integral equation, we generate the perturbative series in the minimal-scheme coupling constants, up to the first order where the $\zeta_3$ appears. To all perturbative orders, at spin one-half, the perturbative series relates to that of the $\beta^2=8\pi^-$ massive sine-Gordon by a sign flip of the coupling constant, and this relation generalizes for an arbitrary spin, to certain massive deformation of $su(2)_{k=2s}$ WZW. This relation supports the conjectured field-theoretical interpretations of such critical points,  at the leading-power $\&\&$ all-logarithmic accuracy. We also compute the perturbative coefficients attached to the first exponentially-small corrections up to the order where $\zeta_3$ appears, in case $s>1$. An interesting feature is that the largest exponentially-small corrections at the locations $h^{2+\frac{2k}{s}}$ are non-vanishing, but the corresponding Borel singularities in the perturbative series are canceled for generic $s>1$. This becomes most dramatic in a 't Hooft type large $s$ limit, in which the perturbative series has a finite convergence radius, while these exponentially small corrections still survive. }

\date{\today}

\begin{document}
\maketitle
\flushbottom

\section{Introduction}
It remains a series of conjectures, that when approaching the CFT-like massless scaling limits, complicated physics becomes simple, and non-perturbative quantities allow perturbative asymptotics, sometimes computed from the text-book style, Feynman-diagram calculations that suffer from various ambiguities/divergences requiring special treatments, that always appear not well-posed enough to those who care about that.  The most mysterious examples of such massless asymptotics are the renormalizable perturbative series generated by {\it marginal operators}: they are more mysterious than those induced by the irrelevant or relevant operators, because a {\it single} non-perturbative parameter like the $\Lambda_{ \overline{\rm MS}}$ locks {\it infinitely-many} computable terms, that really measure the largest deviations to the scale-invariance, that do no really belong to the fixed points.  The ``marginal triviality"~\cite{Aizenman:2019yuo} would be lifted  to the ``marginal-universality", if the conjectures were really correct. 

Unfortunately, in very few cases, one can really test them, especially at high-orders. To the knowledge of the author, there are mainly two exceptions. The first is the large-$N$ expansion of certain 2D QFTs, for which the OPE with condensates can be tested at the level of two-point correlators~\cite{Marino:2024uco,Liu:2025bqq}, currently at NLO in the large-$N$. Another is the free-energy of 2D integrable QFTs in the presence of external fields coupled to conserved charges~\cite{Hasenfratz:1990zz,Hasenfratz:1990ab,Forgacs:1991rs,Zamolodchikov:1995xk}. In this case, it is possible~\cite{Marino:2019eym,Bajnok:2021dri,Marino:2021dzn,Marino:2022ykm,Bajnok:2022xgx,Bajnok:2022rtu,Marino:2023epd,Serone:2024uwz} to generate from the TBA, the exact trans-series expansion at very high perturbative orders, both for the purely perturbative parts and the coefficients attached to the exponentially-small terms. On the side of field-theoretical computations, it is harder to reach similar orders, but there seems to be little doubt that the perturbation theory is correct.

On the other hand, it seems to the author, that the {\it massless-$\phi^4_4$ type} perturbative series near IR fixed-points, induced by marginally-irrelevant operators, were not among the discussions. This note aims to study a class of such examples. To be more precise, we show that the Wiener-Hopf method in Refs.~\cite{Marino:2019eym,Bajnok:2021dri,Marino:2021dzn,Marino:2022ykm,Bajnok:2022xgx,Bajnok:2022rtu,Marino:2023epd} can also be adopted with  minor modifications,  to compute the exact asymptotics of the integrable critical anti-ferromagnetic XXX spin chains~\cite{Babujian:1982ib,Babujian:1983ae} with arbitrary spin $s$, at the level of free-energy in the small external-field limit. For spin-$\frac{1}{2}$, this is the Heisenberg anti-ferromagnetic spin chain. They are all critical, and their IR limits are widely
conjectured to be the $su(2)$ WZW CFTs ~\cite{Witten:1983ar,Knizhnik:1984nr,Affleck:1985jc,Affleck:1988px}.

But beyond that, not too much is known. In fact, although the logarithmic corrections were known to be existing in the early days of QFT~\cite{PhysRev.133.A768}, for a long time, only the first one or two logarithms are computed~\cite{Babujian:1983ae, Granet:2019bum}. It was not checked through serious computations, that such corrections are indeed due to the perturbative series generated by the possible marginal operator $g\bar J^a J^a$, not to mention detailed properties on the exponentially small corrections to them. On the QFT side, the same perturbative series is believed to arise, with a change of sign in the coupling constant, also in certain integrable deformation of $su(2)_k$ WZWs~\cite{Evans:1994hi}, and was computed to high orders in~\cite{Schepers:2023dqk} using the Volin's method~\cite{Volin:2009wr,Volin:2009tqx}. The goal of this work is to show that the same perturbative series with a sign flip indeed arise in the IR limit of the integrable XXX spin chain, but the structure of the exponentially small corrections and their correlation with the perturbative series, are quite different from the QFT case.

It is reasonable to believe that the XXX spin chains are the simplest lattice models supporting the $\phi^4_4$-type perturbative series. We hope this work represents a helpful computation with valuable results.

\section{Analysis of TBA for critical spin chains }\label{sec:spin}
\subsection{Wiener-Hopf analysis of the TBA equation}\label{sec:frame}
We are interested in the ground state energy of the integrable critical anti-ferromagnetic XXX spin chain~\cite{Babujian:1982ib,Babujian:1983ae} with spin $s$ in the small external field $h\rightarrow 0$ limit. It has been shown that the ground state in the absence of a magnetic field is described by a fermi sea of $2s$-strings. The fermi sea exhausted the whole parameter space $-\infty<\lambda<\infty$ and the energy required to create a hole is $\frac{1}{2\cosh \lambda}$. After turning on the external field, a fermi sea consisting of holes in the range $|\lambda|>b$ will be created.  The distribution is determined by the TBA equation~\cite{Babujian:1983ae}
\begin{align}
\epsilon_{2s}(\lambda)+\int_{|\lambda|>b}  J(\lambda-\lambda')\epsilon_{2s}(\lambda')d\lambda'=h-\frac{1}{2\cosh \lambda} \ , \  \lambda>b \ ,
\end{align}
with the boundary condition
\begin{align}
\epsilon_{2s}(b)=0 \ .
\end{align}
The ground state free-energy reads
\begin{align}
F(h)=-\int_{|\lambda|>b}\frac{\epsilon_{2s}(\lambda)}{2\cosh \lambda}\frac{d\lambda}{2\pi} \ .
\end{align}
We now study the above equations in the $h\rightarrow 0$, or equivalently, the $b\rightarrow \infty$ limit. For this purpose, we introduce the function $y(\lambda)=\epsilon_{2s}(\lambda+b)$, then the equation for $y$ reads
\begin{align}
y(\lambda)=h-\frac{1}{2\cosh (\lambda+b)}-\int_{0}^{\infty}J(\lambda-\lambda')y(\lambda')d\lambda'-\int_{0}^{\infty}J(\lambda+\lambda'+2b)y(\lambda') d\lambda' \ ,
\end{align}
subjected to the boundary condition $y(0)=0$. The energy in terms of $y$ reads
\begin{align}
F(h)=-\frac{1}{2\pi} \int_{0}^{\infty} \frac{y(\lambda)}{\cosh(\lambda+b)} d\lambda \ .
\end{align}
We simplify the above equation using the Wiener-Hopf trick. We first extend the integral equation to $\lambda \in (-\infty,\infty)$, with $y(\lambda)=\theta(\lambda)y(\lambda)$ supported in the positive real axis, and a new function $Y(\lambda)$  supported in $(-2b,0)$. We then Fourier transform the integral equation. The convention of the Fourier transform is 
\begin{align}
f(\lambda)=\frac{1}{2\pi}\int_{-\infty}^{\infty}e^{-i\omega\lambda}f(\omega) d\omega \ .
\end{align}
In this convention,  tempered distributions supported in $\lambda \in \mathbb{R}_+$ are transform to analytic functions in the upper half-plane.   Then, after Fourier transform, one has
\begin{align}\label{eq:inteorigin}
\frac{1}{G_+(\omega)G_-(\omega)}y(\omega)+e^{-2i\omega b}J(\omega)y(-\omega)=(h-g)(\omega)+Y(\omega) \ .
\end{align}
Here $(h-g)(\lambda) \equiv (h-g(\lambda))\theta(\lambda)$ with $g(\lambda)=\frac{1}{2\cosh (\lambda+b)}$. The integration kernel has the Wiener-Hopf decomposition
\begin{align}
(1+J)^{-1}(\omega)=G_{+}(\omega)G_-(\omega) \ , \\
G_-(\omega)=
G_+(-\omega)=\sqrt{\frac{\pi}{s}}\frac{\Gamma(\frac{i\omega}{2})}{\Gamma(is\omega)}\frac{e^{is\omega\left(\ln is\omega-1\right) }}{\Gamma(\frac{1}{2}+\frac{i\omega}{2})} \ ,
\end{align}
where $G_+,\   G_+^{-1}$ are analytic and bounded in the upper-half plane, and the same for $G_-, \  G_-^{-}$ in the lower half plane. Their large $\omega$ behaviors in the half-planes with analyticity are of the form $1+{\cal O}(\frac{1}{\omega})$. Furthermore, $G_-(0)=G_+(0)=2s\sqrt{\frac{1}{s}}$ , and $G_-(-i)=G_+(i)=\sqrt{\frac{\pi}{s}}\frac{\Gamma(\frac{1}{2})}{\Gamma(s)}e^{s\ln \frac{s}{e}}$ are non-vanishing.  One can rewrite the Eq.~(\ref{eq:inteorigin}) as
\begin{align}
\frac{1}{G_+(\omega)}y(\omega)+e^{-2i\omega b}G_-(\omega)J(\omega)y(-\omega)=G_-(\omega)(h-g)_+(\omega)+G_-(\omega)Y(\omega) \ .
\end{align}
Let's project it to the upper half plane using the projection 
\begin{align}
f(\omega)_{+}=-\frac{1}{2\pi i}\int_{-\infty}^{\infty} \frac{f(\omega')}{\omega-\omega'+i0}d\omega' \ .
\end{align}
Since $Y(y)$ is compacted supported, $Y(\omega)$ is an entire function and decay as ${\rm Im}(\omega) \rightarrow -\infty$. As such, the last term $G_-Y$  vanishes in this projection. Furthermore, we can write 
\begin{align}
e^{-2i\omega b}G_-(\omega)J(\omega)y(-\omega)=-e^{-2i\omega b}G_-(\omega)y(-\omega)+e^{-2i\omega b}G_-(\omega)(1+J)(\omega)y(-\omega) \ .
\end{align}
Since $y(0)=0$, the large $\omega$ behavior of $y(\omega)$ is ${\cal O}(\omega^{-2})$. On the other hand, the $\omega \rightarrow 0$ singularity of $y(\omega)$ is of the type $(\omega+i0)^{-1}$. As such, the first term on the right side also vanishes under the projection. This leads to the integral equation 
\begin{align}
v(\omega)-\int_{-\infty}^{\infty}\frac{e^{2ib\omega'}}{\omega+\omega'+i0}\frac{G_+(\omega')}{G_-(\omega')}v(\omega')\frac{d\omega'}{2\pi i}=\bigg(G_-(h-g) \bigg)_+ \ ,
\end{align}
for 
\begin{align}
v(\omega)=\frac{y(\omega)}{G_+(\omega)} \ .
\end{align}
We expand $g(\lambda)$ near the end point
\begin{align}
g(\lambda)=\frac{1}{e^{b+\lambda}+e^{-b-\lambda}}=e^{-b-\lambda}\bigg(1+\sum_{n=1}^{\infty}(-1)^ne^{-2n b}e^{-2n \lambda}\bigg) \ .
\end{align}
which implies 
\begin{align}
(G_-g_+)_+=\frac{iG_+(i)e^{-b}}{\omega+i}+\sum_{n=1}^{\infty}(-1)^ne^{-(2n+1)b}\frac{iG_+((2n+1)i)}{\omega+(2n+1)i} \ .
\end{align}
On the other hand, since $G_+(0)=0$, one has
\begin{align}
(G_-h_+)_+=\frac{ihG_+(0)}{\omega+i0} \ .
\end{align}
It is this term that sources the $\omega \rightarrow 0$ singularity of the $v(\omega)$. The $e^{-(2n+1)b}$ terms are exponentially small in the large $b$ limit and are likely due to finite-lattice effects. There is no difficulty to include them in the discussion. They must be added to obtain the ${\cal O}(h^4)$ order corrections, which are the leading corrections at $s=1$ and $s=\frac{1}{2}$. However, since our motivation is to obtain the perturbative parts and the first exponentially small corrections when $s>1$,  we will neglect them in the rest of the work. One then has
\begin{align}
v(\omega)-\int_{-\infty}^{\infty}\frac{e^{2ib\omega'}}{\omega+\omega'+i0}\frac{G_+(\omega')}{G_-(\omega')}v(\omega')\frac{d\omega'}{2\pi i}=\frac{ihG_+(0)}{\omega+i0}-\frac{iG_+(i)e^{-b}}{\omega+i} \ .
\end{align}
From the above, the condition $v(\omega \rightarrow \infty)={\cal O}(\frac{1}{\omega^2})$ (which implies $y(0)=0$) leads to
\begin{align}
ihG_+(0)-ie^{-b}G_+(i)+\frac{1}{2\pi i }\int_{-\infty}^{\infty} d\omega'e^{2ib\omega'} \frac{G_+(\omega')}{G_-(\omega')}v(\omega')=0 \ .
\end{align}
Using this and introducing the function
\begin{align}
u(\omega)=\frac{\omega+i}{ih G_-(0)} v(\omega) \ ,
\end{align}
the integral equation reads
\begin{align}
&u(\omega)=\frac{i}{\omega}-\int_{-\infty}^{\infty} \rho(\omega')\frac{e^{2i\omega' b}}{\omega+\omega'+i0}u(\omega')\frac{d\omega'}{2\pi i} \ , \label{eq:inteu} \\ 
&\rho(\omega)=\frac{G_+(\omega)}{G_-(\omega)}\frac{\omega-i}{\omega+i} \ . 
\end{align}
Comparing with similar equations in integrable QFTs~\cite{Hasenfratz:1990zz,Hasenfratz:1990ab,Forgacs:1991rs,Zamolodchikov:1995xk,Marino:2019eym,Bajnok:2021dri,Marino:2021dzn,Marino:2022ykm,Bajnok:2022xgx,Bajnok:2022rtu,Marino:2023epd}, one can see that the numerator and denorminator are flipped $\rho \rightarrow \rho^{-1}$. In fact, an equivalent integral equation was proposed in~\cite{Marino:2021dzn} by changing $b\rightarrow-b$, $\rho \rightarrow -\rho$ and $i0 \rightarrow -i0$ in the QFT version, as a way to probe the singularities of the integral kernel in the lower-half plane, which move to the upper side in our case. It is interesting to see that this proposal is naturally realized in the spin-chain context. To further simplify, we note that we can also set $\omega$ to be in the lower half-plane in Eq.~(\ref{eq:inteu}), and defines $u(-i)$ in this way. Given this understanding,  the boundary condition can be simplified to
\begin{align}\label{eq:boundary}
G_+(i)e^{-b}=-hG_-(0)u(-i) \ .
\end{align}
Given the above, the free-energy can be written as
\begin{align}
F=-\frac{1}{\pi}\bigg(e^{-b}v(i)G_+(i)+\sum_{n=1}^{\infty}(-1)^ne^{-(2n+1)b}v((2n+1)i)G_+((2n+1)i)\bigg) \ ,
\end{align}
where $v$ is evaluated in the Fourier space. By definition one has
\begin{align}
v(i)=\frac{hG_+(0)}{2}\bigg(1-\int_{-\infty}^{\infty}\frac{d\omega}{2\pi i}\frac{e^{2ib\omega}\rho(\omega)u(\omega)}{\omega+i}\bigg)\equiv \frac{hG_+(0)u(i)}{2} \ .
\end{align}
Combining the above, one has the following compact expression up to power $h^4$
\begin{align}\label{eq:free}
F=\frac{h^2}{2\pi}G_+(0)^2u(-i)u(i)+{\cal O}(h^4)\ .
\end{align}
The ${\cal O}(h^4)$ terms can be added without major modification to the formalism. They can be used to test the irrelevant operators.

\subsection{Singularities of the Wiener-Hopf factors}\label{sec:anal}
We need to consider the Eq.~(\ref{eq:inteu}), Eq.~(\ref{eq:boundary}) and Eq.~(\ref{eq:free}) in the large $b$ limit. It is reasonable to expect that for $\omega$ away from the scaling region $b|\omega|={\cal O}(1)$, the integral in Eq.~(\ref{eq:inteu}) represents a small correction due to the large oscillation of the phase $e^{2ib\omega'}$, and can be treated perturbatively. Near the scaling region, the two terms would be coupled, and it is more involved to determine the $b\rightarrow \infty$ asymptotics in that case. For our model and more general fermionic models~\cite{Marino:2021dzn}, the free energy can be expressed through $u(i)u(-i)$ which is indeed away from the scaling region. In this case there is a simple iterative solution~\cite{Marino:2021dzn} to Eq.~(\ref{eq:inteu}). Below we review how it works in our case, and proceed to the $v^4$ order in the coupling constant expansion, which we believe has its own merits, as all the integrals need to be evaluated exactly in this approach. There is a much more efficient Volin's  method~\cite{Volin:2009wr,Volin:2009tqx}  to obtain the perturbative series at high-orders for TBA in integrable QFT. This method was previously adopted~\cite{Schepers:2023dqk} to certain integrable deformation of $su(2)_k$ WZW~\cite{Evans:1994hi} up to high-orders\footnote{At the time of finishing the first version of this draft, the author was not unaware of the references~\cite{Evans:1994hi,Schepers:2023dqk}.}. We will show later that our purely perturbative part relates to that case by a flip of sign in the coupling constant. But the exponentially small corrections are quite different.

The usual way of performing the large $b$ expansion in such expression is to deform the contour to the upper-half plane and pick up the corresponding singularities. Here the kernel $\rho(\omega)$ has branch-cut singularity starting at $\omega=0$, which can be chosen along the imaginary axis. On top of that, there are two series of poles. The left and right limits of the kernel read
\begin{align}
\rho(i\eta \pm 0)=-\frac{\Gamma(\frac{\eta}{2})}{\Gamma(-\frac{\eta}{2})}\frac{\Gamma(-s\eta)}{\Gamma(s\eta)}\frac{\Gamma(\frac{3-\eta}{2})}{\Gamma (\frac{3+\eta}{2})}e^{2s\eta(\ln s\eta-1)\pm is\pi \eta} \ ,
\end{align}
while the locations of the poles are 
\begin{align}
&\eta_n^A=3+2n, \ n\ge 0 \ , \\ 
&\eta_n^B=\frac{n}{s}, \  n\ge 1 \ .
\end{align}
Notice the second series of poles vanish at $s=\frac{1}{2}$. They will contribute to exponentially-small corrections starting from the order $e^{-6b}\sim h^6$ for the first series,  and $e^{-\frac{2b}{s}}\sim h^{\frac{2}{s}}$ for the second series. For $s>1$ , the latter represents the largest power correction. On the other hand, the discontinuity across the branch cut is given by
\begin{align}
\sigma(\eta)=-\frac{\rho(i\eta+0)-\rho(i\eta-0)}{2\pi i \eta}=\frac{\sin \pi s \eta}{\pi \eta}\frac{\Gamma(\frac{\eta}{2})}{\Gamma(-\frac{\eta}{2})}\frac{\Gamma(-s\eta)}{\Gamma(s\eta)}\frac{\Gamma(\frac{3-\eta}{2})}{\Gamma (\frac{3+\eta}{2})}e^{2s\eta(\ln s\eta-1)} \ .
\end{align}
An interesting feature is that {\it the poles at $s\eta=n$ are canceled in the discontinuity, for generic $s>1$}. From the above, we can introduce the function
\begin{align}
g(\eta)=\eta u(i\eta) \ .
\end{align}
The integral equation up to the first exponentially small correction then reads
\begin{align}
g(\eta)=1+{\rm PV}\int_{0}^{\infty}\frac{\eta }{\eta+\eta'}\sigma(\eta')g(\eta')e^{-2\eta'b}d\eta'+\frac{a(s)\eta e^{-\frac{2b}{s}}}{\eta+\frac{1}{s}}g(\frac{1}{s}) \ ,
\end{align}
where one has
\begin{align}
a(s)=-\frac{\Gamma(\frac{1}{2s})}{\Gamma(-\frac{1}{2s})}\frac{\Gamma(\frac{3s-1}{2s})}{\Gamma(\frac{3s+1}{2s})}e^{-2} \ ,
\end{align}
which is due to the residue at the $\eta s=1$ and is non-vanishing.  In terms of the above, the free-energy reads
\begin{align}
F=-\frac{h^2}{2\pi}G_+(0)^2g(-1)g(1) \ ,
\end{align}
up to power $h^{2+\frac{2}{s}}$.

Here we comment on the following subtlety: {\it  at $s=1$, the pole of $\Gamma(-s\eta)$ collide with the pole at $\eta'=-1$ for $g(-1)$, creating a double pole. 
To treat it properly, one should also include the pole at $\eta'=\eta$
\begin{align}\label{eq:s=1}
&g(-\eta)=1+{\rm PV}\int_{0}^{\infty}\frac{\eta }{\eta-\eta'}\sigma(\eta')g(\eta')e^{-2\eta'b}d\eta'\nonumber \\ 
&+\frac{a(s)\eta e^{-\frac{2b}{s}}}{\eta-\frac{1}{s}}g(\frac{1}{s})+\frac{\rho_+(i\eta)+\rho_-(i\eta)}{2}g(\eta)e^{-2\eta b} \ , 
\end{align}
for $\eta>0$ and then take the $s\rightarrow 1$ limit. This will kill the $\frac{1}{s-1}$ poles as one can check. On the other hand, since for $s=1$, the  $e^{-\frac{2b}{s}}$ terms are at the same order as the omitted $h^4$ terms in Eq.~(\ref{eq:free}), we will not consider them and only present the non-perturbative contribution at $e^{-\frac{2b}{s}}$ when $s>1$. The double pole at $s=1$ will reflects through the $(s-1)^{-1}$ singularity in the expressions. But the purely perturbative parts are free from this issue and the results for them are valid for $s\le 1$ as well. }

\subsection{Perturbative series up to $v^4$ from iterative solution}\label{sec:perturb}
We now rewrite the above as an expansion in terms of a running coupling constant in the minimal scheme. We introduce the coupling in $b$
\begin{align}
\frac{1}{v}+2s\ln \frac{vs2^{\frac{1}{s}}}{e}=2b \ , 
\end{align}
and write
\begin{align}
&\sigma(\eta)=f(\eta)e^{2s\eta(\ln s\eta-1)+2\eta \ln 2} \ , \\ 
&f(\eta)=\frac{\sin \pi s \eta}{\pi \eta}\frac{\Gamma(\frac{\eta}{2})}{\Gamma(-\frac{\eta}{2})}\frac{\Gamma(-s\eta)}{\Gamma(s\eta)}\frac{\Gamma(\frac{3-\eta}{2})}{\Gamma (\frac{3+\eta}{2})}e^{-2\eta \ln 2} \ .
\end{align}
one has 
\begin{align}
g(\eta)=1+v{\rm PV}\int_{0}^{\infty} d\eta' \frac{\eta f(v\eta')}{\eta+v\eta'}e^{-\eta'+2vs\eta' \ln \eta'}g(v\eta')+ \frac{a(s)\eta }{\eta+\frac{1}{s}}g(\frac{1}{s})e^{-\frac{1}{sv}}\left(\frac{e}{vs2^{\frac{1}{s}}}\right)^{2}  \ ,
\end{align}
Clearly, up to power $h^{2+\frac{2}{s}}$, the solution can be solved as
\begin{align}
g(\eta)=g^{(0)}(\eta)+e^{-\frac{1}{sv}}\left(\frac{e}{vs2^{\frac{1}{s}}}\right)^2 a(s)g^{(0)}(\frac{1}{s})g^{(1)}(\eta) \ , \\ 
\end{align}
with
\begin{align}
g^{(0)}(\eta)=1+v{\rm PV}\int_{0}^{\infty} d\eta' \frac{\eta f(v\eta')}{\eta+v\eta'}e^{-\eta'+2vs\eta' \ln \eta'}g^{(0)}(v\eta') \ , \label{eq:g0}
\end{align}
and
\begin{align}
g^{(1)}(\eta)=\frac{\eta}{\eta+\frac{1}{s}}+v{\rm PV}\int_{0}^{\infty} d\eta' \frac{\eta f(v\eta')}{\eta+v\eta'}e^{-\eta'+2vs\eta' \ln \eta'}g^{(1)}(v\eta')  \ .
\end{align}
One can actually include higher power corrections in a more systematic way as in Refs.~\cite{Marino:2021dzn,Bajnok:2022xgx,Marino:2023epd} to obtain the expansion terms at high orders. Up to this stage, no perturbative expansion has been performed in $g^{(0)}$ and $g^{(1)}$.  They can be further expanded in  $v$ by iterating the integral equations like in~\cite{Marino:2021dzn}. More explicitly, one has
\begin{align}
g^{0}(\eta)=1+\sum_{n=1}^{\infty}v^n\int_{0}^{\infty}d\eta_1...d\eta_n\frac{\eta \eta_1..\eta_{n-1}f(v\eta_1)..f(v\eta_{n})}{(\eta+v\eta_1)(\eta_1+\eta_2)..(\eta_{n-1}+\eta_n)}e^{-\sum_{i=1}^{n}(\eta_i-2vs\eta_i\ln \eta_i)} \ , 
\end{align}
and
\begin{align}
&g^{(1)}(\eta)=\frac{\eta}{\eta+\frac{1}{s}}\nonumber \\ 
&+\sum_{n=1}^{\infty}v^{n+1}\int_{0}^{\infty}d\eta_1...d\eta_n\frac{\eta \eta_1..\eta_{n-1}f(v\eta_1)..f(v\eta_{n})}{(\eta+v\eta_1)(\eta_1+\eta_2)..(\eta_{n-1}+\eta_n)}\frac{\eta_n}{\frac{1}{s}+v\eta_n}e^{-\sum_{i=1}^{n}(\eta_i-2vs\eta_i\ln \eta_i)} \ .
\end{align}
Expanding the integrands in $v$, all the coefficients are represented as absolutely convergent integrals.  In particular, one can expand $g^{(0)}$ and $g^{(1)}$ to $v^4$ with a moderate amount of workload.  For this one needs the integrals
\begin{align}
\int_{0}^{\infty}d\eta_1d\eta_2d\eta_3\frac{\eta_1\eta_2}{(\eta_1+\eta_2)(\eta_2+\eta_3)}e^{-\eta_1-\eta_2-\eta_3}=\frac{7}{2}-\frac{\pi^2}{3} \ , \\
\int_{0}^{\infty}d\eta_1d\eta_2d\eta_3d\eta_4\frac{\eta_1\eta_2\eta_3}{(\eta_1+\eta_2)(\eta_2+\eta_3)(\eta_3+\eta_4)}e^{-\eta_1-\eta_2-\eta_3-\eta_4}=\frac{27}{8}-\frac{\pi ^2}{3} \ .
\end{align}
They are similar to massive Feynman integrals of vacuum diagrams, in the Schwinger parameter space. Some of the integrals with 5 and 6 parameters are still manageable using the Mellin-Barnes method, for example one has 
\begin{align}
&\int_0^{\infty}\prod_{i=1}^5d\eta_i \frac{\eta_1\eta_2 \eta_3\eta_4e^{-\sum_{l=1}^5\eta_l}}{\prod_{j=1}^4(\eta_j+\eta_{j+1})}=\frac{517}{24}-\frac{359}{18}\pi^2+\frac{9}{5}\pi^4 \ , \\
&\int_0^{\infty}\prod_{i=1}^6d\eta_i \frac{\eta_1\eta_2 \eta_3\eta_4\eta_5e^{-\sum_{l=1}^6\eta_l}}{\prod_{j=1}^5(\eta_j+\eta_{j+1})}=\frac{1247}{48}-\frac{377}{18}\pi^2+\frac{167}{90}\pi^4 \ . 
\end{align}
Let's mention that the difficulty of these integrals lies not just in the increased parameter numbers. Even with three parameters, when the $\ln \eta_i$s start to accumulate after expanding the exponentials, the integrals become as challenging as the ones with higher parameter numbers,  but fewer or no logarithms. This is similar to what happened, when trying to expand the dimensional regularized Feynman integrals that are not gamma and polygamma ratios, to high orders in $\epsilon$.  It happens that in the final results at $v^5$ and $v^6$, all the $\pi^2$s and $\pi^4$s are canceled by other integrals with fewer parameters. It is reasonable to expect that at even higher orders, not all the individual integrals can be computed explicitly into $\zeta$ values, but the total combination does. The Volin's method~\cite{Volin:2009wr,Volin:2009tqx} works much more effective in the generic cases. 

After many cancellations among the integrals, the full result ($s\ne1$) to power $e^{-\frac{1}{sv}}$ reads
\begin{align}\label{eq:Free}
F(h,s)=-\frac{h^2}{2\pi}G_+(0)^2\bigg(f^{(0)}(v,s)+e^{-\frac{1}{sv}}\left(\frac{e}{vs2^{\frac{1}{s}}}\right)^2 a(s)f^{(1)}(v,s)\bigg) \ ,
\end{align}
where one has
\begin{align}
f^{(0)}(v,s)=&1+2sv+2 \left(-2s+3 s^2\right) v^2+4 \left(3s-8 s^2+6s^3\right) v^3\nonumber \\ 
&+2 (-24 s + 81 s^2 - 104 s^3 + 53 s^4 + 3 s \zeta_3) v^4+{\cal O}(v^5) \ , \\ 
f^{(1)}(v,s)=&\frac{2  s^2}{(s-1) (s+1)}+\frac{4  s^3 v}{(s-1) (s+1)}\nonumber \\ 
&+\frac{4  \left(s^2+3 s^3\right) v^2}{s+1}+\frac{4 \left(12 s^2-6s^3-48s^4+37s^5\right) v^3}{3 (s-1) (s+1)}\nonumber \\
&+\frac{2 \left(-108 s^2+192 s^3+186 s^4-636 s^5+331 s^6+18 s^3\zeta_3\right)v^4}{3 (s-1) (s+1)}+{\cal O}(v^5) \ . \label{eq:f1}
\end{align}
Notice that the coefficients are all rational up to ${\cal O}(v^3)$. Eq.~(\ref{eq:Free}) is the major result of this paper. Finally,  introducing the running coupling constant of $h$ in the minimal scheme
\begin{align}
\ln \frac{G_+(i)}{G_-(0)h}=\frac{1}{2\alpha}+s\ln \frac{\alpha s2^{\frac{1}{s}}}{e} \ ,
\end{align}
the boundary condition Eq.~(\ref{eq:boundary}) can be written as
\begin{align}
e^{\frac{1}{2\alpha}-\frac{1}{2v}+s\ln\frac{\alpha}{v}}=g^{(0)}(-1)+e^{-\frac{1}{vs}}\left(\frac{e}{vs2^{\frac{1}{s}}}\right)^2 a(s)g^{(0)}(\frac{1}{s})g^{(1)}(-1)  \ ,
\end{align}
where
\begin{align}
g^{(0)}(-1)&=1+sv+\frac{1}{2} \left(-2s+5 s^2\right) v^2+\frac{1}{6} \left(12 s-44s^2+57s^3\right) v^3\nonumber \\ 
&+\frac{1}{24}(-144 s+612 s^2-1088 s^3+969 s^4+72 s \zeta_3) v^4+{\cal O}(v^5) \ , \label{eq:g0mius} \\ 
g^{(0)}(\frac{1}{s})g^{(1)}(-1)&=\frac{s}{s-1}+\frac{s^2 v}{s-1}+\frac{\left(-6 s^2+5s^3\right) v^2}{2 (s-1)}+\frac{\left(36 s^2-104s^3+61s^4\right) v^3}{6 (s-1)} \nonumber \\ 
&+\frac{(-432 s^2+1572 s^3-2312 s^4+1057 s^5+72 s^2 \zeta_3) v^4}{24(s-1)}+{\cal O}(v^5) \ .
\end{align}
The above allows the conversion from $v$ to the ``minimal scheme'' $\alpha$ as $v=\alpha v_0(\alpha)+e^{-\frac{1}{s\alpha}}\alpha^{-2}v_1(\alpha)$ where $v_0(\alpha)$ and $v_1(\alpha)$ are regular formal power series. In the following, we use mainly the coupling $v$. 
\subsection{The special case $s=\frac{1}{2}$ and relation to $\beta^2 \rightarrow 8\pi^-$ sine Gordon}\label{sec:spinhapf}
It is interesting to notice that in the spin-$\frac{1}{2}$ 
case, the $G_{\pm}(\omega)$ is identical to the $\beta^2 \rightarrow 8\pi^-$ limit of massive sine Gordon theory, namely 
\begin{align}
\frac{G^{+}(i\eta )}{G^{-}(i\eta \pm 0)}=\frac{\Gamma(\frac{1-\eta}{2})}{\Gamma(\frac{1+\eta}{2})}e^{\eta\left(\ln \frac{\eta}{2}-1\right)}e^{\pm \frac{i\pi \eta}{2}} \ .
\end{align}
This leads to the equation for the PT part
\begin{align}
g(\eta)=1+\int_{0}^{\infty} d\eta' e^{-2b\eta'}\frac{\eta}{\eta+\eta'}\frac{\Gamma(\frac{3-\eta'}{2})}{\Gamma(\frac{3+\eta'}{2})}e^{\eta'\left(\ln \frac{\eta'}{2}-1\right)}\frac{\sin \frac{\pi \eta'}{2}}{\pi \eta'}g(\eta')\ .
\end{align}
Using the coupling
\begin{align}
\frac{1}{v}+\ln \frac{2v}{e}=2b \ ,
\end{align}
then the perturbative expansion in this $v$ is formally the same as that for the $\beta^2=8\pi^-$ bootstrap theory with $v \rightarrow -v$. The details for the expansion in the O(2) theory are presented in Appendix.\ref{Sec:O(2)}.  More precisely, for both of the models the free energy is represented as
\begin{align}
F(v)=-\frac{h^2}{\pi}g(-1,v)g(1,v) \ .
\end{align}
The perturbative series satisfies the relation
\begin{align}
g^{\rm s=\frac{1}{2}}(-1,v)=g^{\rm O(2)}(1,-v) \ , \\ 
g^{\rm s=\frac{1}{2}}(1,v)=g^{\rm O(2)}(-1,-v) \ .
\end{align}
This is consistent with the fact that the two cases correspond to the $\pm g \bar J^a J^a$ perturbation to $c=1$ SU(2) WZW~\cite{Witten:1983ar,Knizhnik:1984nr,Affleck:1985jc,Affleck:1988px}. 
\subsection{General $s$: relation to massive deformation of $su(2)_{k=2s}$ WZW}\label{sec:generals}
The relationship to the large $h$ limit of integrable QFT extends to general $s$ as well: the perturbative series for spin $s$ relates to the large $h$ limit of the massive deformation of $su(2)_k$ WZW in~\cite{Evans:1994hi} simply by a sign flip of the coupling constant. The perturbative series and the Borel singularities on the QFT side was previously investigated in~\cite{Schepers:2023dqk}. In fact, in these cases the integral kernels for the corresponding TBA equations are the same, the only difference is the integration range as well as the dispersion relations: 
\begin{align}
&\epsilon_{UV}(\theta)+\int_{|\theta|<b}J(\theta-\theta')\epsilon_{UV}(\theta') d\theta' =h-m\cosh \theta \ , \label{eq:TBAIR} \\
&\epsilon_{IR}(\lambda)+\int_{|\lambda|>b}J(\lambda-\lambda')\epsilon_{IR}(\lambda')d\lambda'=h-e^{-|\lambda|}\ . \label{eq:TBAUV}
\end{align}
Translating into the $u$ functions, it is easy to show that the numerator and denominators flip
\begin{align}
&u(\omega)=\frac{i}{\omega}-\int_{-\infty}^{\infty} \rho(\omega')\frac{e^{2i\omega' b}}{\omega+\omega'+i0}u(\omega')\frac{d\omega'}{2\pi i} \ , \\ 
&\rho_{IR}(\omega)=\frac{G_+(\omega)}{G_-(\omega)}\frac{\omega-i}{\omega+i} \ , \ \rho_{UV}(\omega)=\frac{G_-(\omega)}{G_+(\omega)}\frac{\omega+i}{\omega-i} \ .
\end{align}
As a result, one has 
\begin{align}
\rho_{IR}(i\eta \pm 0)=-\frac{\Gamma(\frac{\eta}{2})}{\Gamma(-\frac{\eta}{2})}\frac{\Gamma(-s\eta)}{\Gamma(s\eta)}\frac{\Gamma(\frac{3-\eta}{2})}{\Gamma (\frac{3+\eta}{2})}e^{2s\eta(\ln s\eta-1)\pm is\pi \eta} \ , \\ 
\rho_{UV}(i\eta \pm 0)=-\frac{\Gamma(-\frac{\eta}{2})}{\Gamma(\frac{\eta}{2})}\frac{\Gamma(s\eta)}{\Gamma(-s\eta)}\frac{\Gamma(\frac{3+\eta}{2})}{\Gamma (\frac{3-\eta}{2})}e^{-2s\eta(\ln s\eta-1)\mp is\pi \eta} \ . \label{eq:rhoUV}
\end{align}
This implies that when one closes the contours around the positive imaginary axis, the discontinuities relate to each other through $\eta \rightarrow -\eta$. Given this, after introducing corresponding coupling constants, it is not hard to see that the perturbative parts are controlled by the following equations of the $g$ functions
\begin{align}
g_{IR}(\eta)=1+v{\rm PV}\int_{0}^{\infty} d\eta' \frac{\eta f(v\eta')}{\eta+v\eta'}e^{-\eta'+2vs\eta' \ln \eta'}g_{IR}(v\eta') \ ,  \\
g_{UV}(\eta)=1-v{\rm PV}\int_{0}^{\infty} d\eta' \frac{\eta f(-v\eta')}{\eta+v\eta'}e^{-\eta'-2vs\eta' \ln \eta'}g_{UV}(v\eta') \ . 
\end{align}
The above implies at the level of the iterative solutions, one has
\begin{align}
g_{UV}(\eta,v)=g_{IR}(-\eta,-v) \ , \\ 
g_{UV}(-\eta,v)=g_{IR}(\eta,-v) \ .
\end{align}
The above implies
\begin{align}
&F_{UV}(v)=-\frac{h^2}{2\pi}G_+^2(0)g_{UV}(1,v)g_{UV}(-1,v) \ , \\ 
&F_{IR}(v)=-\frac{h^2}{2\pi}G_+^2(0)g_{IR}(1,v)g_{IR}(-1,v)\equiv F_{UV}(-v) \ , 
\end{align}
extending the relationship for $s=\frac{1}{2}$. The above applies to an arbitrary pair of TBA equations of the form Eq.~(\ref{eq:TBAIR}), Eq.~(\ref{eq:TBAUV}), controlled by a ``fermionic'' integration kernel $J(\theta-\theta')$. On the other hand, from Eq.~(\ref{eq:rhoUV}), the structure of exponentially small corrections in the $h \rightarrow \infty$ limit of massive QFT is considerably simpler than the marginally irrelevant case of integrable spin chains: when $2s\in \mathbb{Z}_{\ge 1}$, the singularity at $\eta=2n$ is always canceled by the $\Gamma(-s\eta)$ in the denominator (this fact was observed first in~\cite{Schepers:2023dqk} and was related to the ``Cheshire Cat resurgence''~\cite{Kozcaz:2016wvy} therein). This cancellation is 
at the level of the Wiener-Hopf factor and is before taking the discontinuity of $\rho(i\eta)$. As a result,  the would be exponentially-small corrections are also canceled. On the other hand, the singularities on the lower-half plane, called UV renormalons in~\cite{Marino:2021dzn}, exactly move to the upper side in the IR expansion of the spin-chains. In this case, the leading singularities are canceled in the discontinuity of $\rho(i\eta)$ for generic  $s>1$, but not in the residues, and the resulting exponentially small corrections, known to be unambiguous in~\cite{Schepers:2023dqk}, become not detectable in the perturbative series.  The $s$ parametrizes a one-dimensional family of theories in which the Cheshire Cat disappears~\cite{Kozcaz:2016wvy}, on the spin-chain side.

\subsection{A large $s$ expansion}
To magnify the phenomenon mentioned in the last subsection, let's consider the following large-$s$ 't Hooft-type limit
\begin{align}
s \rightarrow \infty, \  \lambda=vs={\cal O}(1) \ . 
\end{align}
This limit requires $h\rightarrow 0$ exponentially fast in $s$. In this limit,  both the perturbative series and the non-perturbative corrections due to the $\eta s=n$ poles survive and remain non-trivial. Unlike the large-$N$ expansion in the models with scalar interactions such as the Gross-Neveu or NLSM, the large $s$ expansion here seems more like the YM version, and the iterative solution does not truncate. For example, for the $g^{0}(\eta)$ function in Eq.~(\ref{eq:g0}) generating the perturbative series,  the iterative solution at the leading order in the large-$s$ expansion reads 
\begin{align}
& g^{(0)}(\eta)=1+\sum_{n=1}^{\infty} \lambda^n \int_{0}^{\infty}d\eta_1...d\eta_n\frac{\eta_1..\eta_{n-1}\hat f(\lambda \eta_1)..\hat f(\lambda \eta_{n})}{(\eta_1+\eta_2)..(\eta_{n-1}+\eta_n)}e^{-\sum_{i=1}^{n}(\eta_i-2\lambda \eta_i\ln \eta_i)} \ , \\ 
& \hat f( \eta)=-\frac{\sin \pi  \eta}{\pi  \eta}\frac{\Gamma(- \eta)}{\Gamma(\eta)} =\frac{1}{\Gamma^2(1+\eta) } \ .
\end{align}
Introducing the following integral kernel
\begin{align}
{\cal K}(\eta,\eta')=\frac{\eta}{\eta+\eta'}E(\eta)E(\eta')\ , \  E(\eta)=\frac{e^{-\frac{\eta}{2}+\lambda \eta \ln \eta}} {\Gamma(1+\lambda \eta)} \ , 
\end{align}
the iterative solution can be expressed through its resolvent as  
\begin{align}
g^{(0)}(\eta)=1+\bigg\langle E\bigg|\frac{\lambda }{1-\lambda{\cal K}}\bigg|E\bigg\rangle \ . 
\end{align}
Not only the $\eta=k$ poles of $\hat f(\eta)$ are all canceled by the factor $ \sin \pi  \eta $ due to the phase discontinuity across the branch cut, but all the other $\Gamma$ factors also disappear. To see the convergence property of the perturbative series in such case, let's note that for $\eta>0$, we have the following exponential bound
\begin{align}
|E(\eta)|\le C e^{-\frac{\eta}{2}}\big|e^{-\eta \lambda (\ln \lambda -1) }\big| \ , 
\end{align}
uniformly in $\lambda$. Then, for sufficiently small $\lambda$, ${\cal K}$ is of trace class, and $g^{(0)}$ is analytic in $\lambda$. This implies its perturbative series in $\lambda $ has a finite radius of convergence. However, the exponentially-small corrections still exist in the large-$s$ limit, which can be seen from the Eq.~(\ref{eq:f1}).

\section{Summary and outlook}\label{sec:conc}
In this paper, inspired by recent works~\cite{Marino:2019eym,Bajnok:2021dri,Marino:2021dzn,Marino:2022ykm,Bajnok:2022xgx,Bajnok:2022rtu,Marino:2023epd} on trans-series expansion in 2D integrable QFTs, we study similar asymptotic expansions of integrable anti-ferromagnetic spin chains near the IR fixed point, in the $h\rightarrow 0$ limit. The perturbative series can be generated through the iterative solution to the TBA equation.  At any given order, the coefficients are given by a finite number of absolute convergent integrals similar to the Feynman integrals in the $\alpha$-parameter space.
The perturbative series are exactly related to certain massive deformation of $su(2)_k$ WZW. The singularities on the positive Borel axis are  given by the {\rm UV  renormalons}~\cite{Marino:2021dzn} of the latter. For $s>1$, we found that the leading exponentially-small corrections are of the order $h^{2+\frac{2}{s}}$. The would-be Borel poles in the perturbative series are coincidentally canceled. There is a YM-like large-$s$ expansion, in which the perturbative series has a finite radius of convergence, but such exponentially small contributions still survive. 

\acknowledgments
The author thanks István Vona for checking the iterative computation using the Volin's method and sharing the corresponding codes. The author thanks Romuald Janik for pointing out the Ref.~\cite{Evans:1994hi}. Y. L. is supported by the Priority Research Area SciMat and DigiWorlds under the program Excellence Initiative - Research University at the Jagiellonian University in Krak\'{o}w.

\appendix
\section{Perturbative series of the $\beta^2 \rightarrow 8\pi^-$ sine Gordon}\label{Sec:O(2)}
In this appendix, we study the bootstrap theory based on the $\beta^2 \rightarrow 8\pi^-$ of the massive sine Gordon's S-matrix, also called the $O(2)$ bootstrap theory~\cite{Balog:2001wv}. This theory can be identified as the $T\rightarrow T_c^+$ massive scaling limit of the 2D XY model~\cite{Balog:2001wv}. On the other hand, the $O(4)$ Gross-Neveu model, at least at the level of $S$ matrix, decouples into two identical copies of this theory~\cite{Witten:1977xv}.  As usual~\cite{Hasenfratz:1990zz,Hasenfratz:1990ab,Forgacs:1991rs,Zamolodchikov:1995xk,Marino:2019eym,Bajnok:2021dri,Marino:2021dzn,Marino:2022ykm,Bajnok:2022xgx,Bajnok:2022rtu,Marino:2023epd}, one consider the ground state energy $F(h)$ in the presence of a back ground field $h$ and study $h\gg m$ limit. Compared with the $O(N)$ ($N\ge 3$) case analyzed in these references, the analysis of $O(2)$ is much simpler due to the fact that the $K_-(\omega)$ has only branch cut singularity in the upper-half plan but no pole, and $K_{\pm}(0)$ is finite, suggesting perturbative dominance. The result can be summarized as follows, see~\cite{Zamolodchikov:1995xk,Samaj:2013yva,Reis:2022tni} for detailed derivation of integral equations.

First, one needs the Wiener-Hopf decomposition $K_{\pm}(\omega)$, which is analytic and non-vanishing in the upper (lower) half-plane
\begin{align}
K_{-}(\omega)=\sqrt{2\pi}\frac{e^{\frac{i\omega}{2}\left(\ln \frac{i\omega}{2}-1\right)}}{\Gamma \left(\frac{1}{2}+\frac{i\omega}{2}\right)} \ , \ K_{+}(\omega)=K_-(-\omega) \ .
\end{align}
In terms of them, one needs the function 
\begin{align}
\rho(\omega)=-\frac{w+i}{w-i}\frac{K_-(\omega)}{K_{+}(\omega)} \ .
\end{align}
It is easy to see that $\rho(w)$ has only branch cut singularity in the upper half plane in the positive imaginary axis but no poles. Moreover, $\rho(\omega) \rightarrow 1+{\cal O}(\frac{1}{\omega})$ everywhere else in the upper half plane when $\omega \rightarrow \infty$. Its discontinuity in the positive imaginary axis reads
\begin{align}
\sigma(\eta)=-\frac{\rho(i\eta+0)-\rho(i\eta-0)}{2\pi \eta i}=\frac{1}{\pi\eta}\sin \frac{\pi \eta}{2}\frac{1+\eta}{1-\eta}\frac{\Gamma \left(\frac{1+\eta}{2}\right)}{\Gamma \left(\frac{1-\eta}{2}\right)}e^{-\eta\left(\ln \frac{\eta}{2}-1\right)} \ .
\end{align}
It is easy to see that $\sigma(\eta)$ is regular for $\eta>0$, $|\sigma(\eta)|<2$ everywhere, and $\sigma(\eta)\rightarrow -\frac{\sin \pi \eta}{\pi \eta}$ as $\eta \rightarrow \infty$.
One now introduces a crucial analytic function determined by the integral equation
\begin{align}
    u(w)=\frac{i}{\omega}+\frac{1}{2\pi i}\int_{\rm R} \frac{e^{2iB\omega'}\rho(\omega')}{\omega+\omega'+i0}u(\omega')d\omega' \ .
\end{align}
In this equation, $B\gg 1$ is a positive number which determines the fermi-energy $E_F=m\cosh B$ of the ground state. In this case, $u(\omega)$ is analytic in the upper half plane as well, which means one can deform the integration path from the real axis to a path surrounding the imaginary axis and pick up the discontinuity $\sigma(\eta)$. This leads to a closed equation for $g(\eta) \equiv \eta u(i\eta)$ ($\eta>0$)
\begin{align}
g(\eta)=1-\int_{0}^{\infty}e^{-2B\eta'} \frac{\eta}{\eta+\eta'}\sigma(\eta')g(\eta')d\eta' \ .
\end{align}
The free energy reads
\begin{align}
F=-\frac{h^2}{\pi}g(1)g(-1) \ .
\end{align}
Here $g(-1)$ should be understood as $u(\omega)$ analytically continued to $-i$ using the integral representation. It t will create renormalon ambiguities as we will see below.
The $B$ relates to $\frac{h}{m}$ through
\begin{align}\label{eq:contition}
g(1)=\frac{me^{B}}{h}\frac{\sqrt{\pi}e^{-\frac{1}{2}}}{2\sqrt{2}} \ .
\end{align}
One now analyse the above equation using the route in the literature. First, the equation for $g(\eta)$ can be iterated as
\begin{align}
g(\eta)=1+\sum_{n=1}^{\infty}(-1)^n\int_0^{\infty}d\eta_1..d\eta_n\frac{\eta \eta_1....\eta_{n-1}\sigma(\eta_1)...\sigma(\eta_n)}{(\eta+\eta_1)..(\eta_{n-1}+\eta_n)}e^{-2B(\eta_1+...\eta_n)} \ .
\end{align}
Using the fact that $|\sigma(\eta)|<1$, the above expansion converges absolutely as $2B>2$, or $B>1$, $E_F\ge 1.54m$. This bound can be improved, but we will not need this. Now, consider
\begin{align}
g_n(\eta)=\int_0^{\infty}d\eta_1..d\eta_n\frac{\eta \eta_1....\eta_{n-1}\sigma(\eta_1)...\sigma(\eta_n)}{(\eta+\eta_1)..(\eta_{n-1}+\eta_n)}e^{-2B(\eta_1+...\eta_n)} \ ,
\end{align}
one can introduce a running coupling $v$ in $B$ to express $g_n$ as a re-summed perturbative expansion. Naively, one simply chose $v=\frac{1}{2B}$, but due to the logarithm in the exponential of $\sigma(\eta)$, expansion in this $v$ will contain logarithms. Write
\begin{align}
\sigma(\eta)=e^{-\eta \left(\ln \frac{\eta}{2}-1\right)-2\eta \ln 2}f(\eta) \ ,
\end{align}
where $f(\eta)$ has regular Taylor expansion at $\eta=0$, it is convenient to introduce the coupling in the following way
\begin{align}
2B=\frac{1}{v}-\ln\frac{2v}{e} \ .
\end{align}
The condition $2B \ge 2$ translates to $v\le 0.727$. In terms of this coupling constant, after the rescaling $\eta_i=v \eta_i$, one has
\begin{align}
g_n(\eta,v)=v^nP_n(v,\eta)=v^n\int_0^{\infty}d\eta_1..d\eta_n\frac{\eta \eta_1....\eta_{n-1}f(v\eta_1)...f(v\eta_n)}{(\eta+v\eta_1)..(\eta_{n-1}+\eta_n)}e^{-(\eta_1+...\eta_n)-v\sum_{i=1}^n\eta_i\ln \eta_i} \ .
\end{align}
Clearly, after expanding the integrand in $v$ now, it becomes a perturbative series in $v$ of the form $v^n(1+g_n^1v+g_n^2v^2+...)$. As a result, one has
\begin{align}
g(v,\eta)=1+\sum_{n=1}^{\infty}(-1)^nv^n P_n(v,\eta) \ , \\ 
P_n(v,\eta) \sim \sum_{k=0}^{\infty}g_n^k(\eta)v^k \ .
\end{align}
Notice that the perturbative series for $P_n(v,\eta)$ is divergent but Borel-summable in terms of our integral representation. The expansion of $g(v,\eta)$ in terms of resummed $P_n(v,\eta)$ is absolutely convergent for  $v \le 1.45$. This can be regarded as a convergent partially resumed perturbative expansion for $g(\eta,v)$. 

We then move to the free energy. It is determined as
\begin{align}
F=-\frac{h^2}{\pi}g(1)\bigg(1-\frac{1}{2\pi i}\int_{\rm R}\frac{e^{2iB\omega'}\rho(\omega')u(\omega')}{\omega'-i}d\omega'\bigg) \ .
\end{align}
Unlike the equation for $u$, the pole at $\omega'=i$ will not be canceled, as a result, one has renormalon ambiguity now. By closing the contour as before, but using small circles to circumvent the singularity, one has
\begin{align}
F=-\frac{h^2}{\pi}g(1)\bigg(1-v{\rm PV}\int_{0}^{\infty}\frac{e^{-\eta-v\eta \ln \eta}f(v\eta)g(v\eta)}{1-v\eta}d\eta\bigg) \ ,
\end{align}
where the pole at $\eta=\frac{1}{v}$ is circumvented using the Principle value prescription. If one use the $\pm i0$ prescription, there will be the ambiguity of order
\begin{align}\label{eq:ambiguity}
\delta F=2\pi i \frac{h^2}{\pi}ve^{-\frac{1}{v}}f(1)g^2(1) \ .
\end{align}
On the other hand, there is no other ambiguity of the free energy anymore. In fact, one can write
\begin{align}
&F=-\frac{h^2}{\pi}g(1)+\frac{h^2}{\pi}g(1)v \sum_{n=0}^{\infty}(-1)^nF_n(v) \ , \\ 
&F_n(v)=v^n {\rm PV} \int_0^{\infty}d\eta_0 d\eta_1..d\eta_n\frac{\eta_0 \eta_1....\eta_{n-1}f(v\eta_0)f(v\eta_1)...f(v\eta_n)}{(1-v\eta_0)(\eta_0+\eta_1)..(\eta_{n-1}+\eta_n)}e^{-\sum_{i=0}^n \eta_i(1+v\ln \eta_i)} \ .
\end{align}
Clearly, when expanded in $v$, the $F_n(v)$ has quadratic renormalon ambiguity. On the other hand, when resummed into $F_n(v)$ using principle value prescription, the expansion above is absolutely convergent for small $v$.

In fact, using the formula which is valid for $\eta>0$
\begin{align}
e^{-\frac{\eta}{v}}v^\eta=\frac{1}{\Gamma(\eta)}\int^{\infty}_{\eta} dt e^{-\frac{t}{v}}\frac{1}{(t-\eta)^{1-\eta}} \ ,
\end{align}
one can show that the leading $t\rightarrow 1$ singularity is purely contrlled by the $\frac{h^2}{\pi}vF_0(v)$ terms. The Borel transform of this term can be represented by the integral 
\begin{align}
B_0(t)=\frac{h^2}{\pi} \int_{0}^t \frac{f(\eta)}{(1-\eta)}\frac{e^{-\eta\ln \eta }}{(t-\eta)^{1-\eta}}\frac{1}{\Gamma(\eta)}d\eta \ .
\end{align}
The singularity as $t\rightarrow 1^-$ reads
\begin{align}\label{eq:leadingborel}
B_0(t) \rightarrow \frac{h^2}{\pi} \times f(1)\ln(1-t) \ .
\end{align}
This leads to the ambiguity of the form
\begin{align}\label{eq:Borelsingularity}
\delta F=\frac{h^2}{\pi^2}f(1)2{\rm Im}\int_{1}^{\infty} dt e^{-\frac{t}{v}}\ln (1-t+i0)=2i\pi \times \frac{h^2}{\pi} f(1)ve^{-\frac{1}{v}} \ ,
\end{align}
which is consistent with the result Eq.~(\ref{eq:ambiguity}). High-order Borel asymptotics neglected here will be required to maintain scheme independence. However, their forms are almost fully determined by this requirement as far as one knows the conversion from $v$ to physical scales in Eq.~(\ref{eq:contition}). In fact, Eq.~(\ref{eq:contition}) can be interpreted as defining the $\beta$ function for the coupling $v$.

To summarize, this simple example shows that for the $\beta^2=8\pi^-$ sine Gordon, although the free energy when expanded in terms of $v$ or $\alpha$ has renormalon ambiguity, the full result is recovered from the perturbative series after using the ``principle value prescription''. Moreover, the ambiguity must be exactly reproduced in the perturbative calculation. It is a good exercise to check this in the fermionic representation~\cite{Witten:1983ar,Knizhnik:1984nr} with ``isospin'' interaction $g^2\bar \psi\gamma^{\mu}\tau^a\psi \bar \psi \gamma_{\mu}\tau_a \psi$ for two flavors of Dirac fermions. It might be possible to implement the dimensional regularization consistently in this representation, without breaking the $SU(2)$ symmetry. However,  due to the presence of a Fermi sea, the integrals start to look not well posed enough again, at least to the author, than those in the Bosonic models. 
\bibliographystyle{apsrev4-1} 
\bibliography{bibliography}

\end{document}